\documentstyle[sprocl,epsfig]{article}
\newcommand{\lesssim}{\raisebox{0.3mm}{\em $\, <$} \hspace{-2.8mm}
\raisebox{-1.3mm}{\em $\sim \,$}}
\newcommand{\gtrsim}{\raisebox{0.3mm}{\em $\, >$} \hspace{-2.8mm}
\raisebox{-1.3mm}{\em $\sim \,$}}

\def\be{\begin{equation}}
\def\ee{\end{equation}}
\def\bea{\begin{eqnarray}}
\def\eea{\end{eqnarray}}

\begin{document}

\title{NEUTRINO OSCILLATIONS WITH FOUR GENERATIONS}

\author{OSAMU YASUDA}

\address{Department of Physics,
Tokyo Metropolitan University \\
Minami-Osawa, Hachioji, Tokyo 192-0397, Japan
\\E-mail: yasuda@phys.metro-u.ac.jp}


\maketitle\abstracts{Recent status of neutrino oscillation
phenomenology with four neutrinos is reviewed.
It is emphasized that the so-called (2+2)-scheme as well as
the (3+1)-scheme are still
consistent with the recent solar and atmospheric neutrino data.}

\section{Introduction}

There have been several experiments
\cite{homestake,Kamsol,SKsol,sage,gallex,Kamatm,IMB,SKatm,ysuzuki}
\cite{SKs,soudan2,lsnd,mills}
which suggest neutrino oscillations.
To explain the solar, atmospheric and LSND data within the framework
of neutrino oscillations, it is necessary to have at least four kinds of
neutrinos.
It has been shown in the two
flavor framework that the solar neutrino deficit can be explained by
neutrino oscillation with the sets of parameters $(\Delta
m^2_\odot,\sin^22\theta_\odot)\simeq$ $({\cal O}(10^{-5}{\rm
eV}^2),{\cal O}(10^{-2}))$ (SMA (small mixing angle) MSW solution),
$({\cal O}(10^{-5}{\rm eV}^2),{\cal O}(1))$ (LMA (large mixing angle)
MSW solution), $({\cal O}(10^{-7}{\rm eV}^2),{\cal O}(1))$
(LOW solution)
or $({\cal O}(10^{-10}{\rm eV}^2),{\cal O}(1))$ (VO
(vacuum oscillation) solution).
At the Neutrino 2000 Conference the Superkamiokande group
has updated their data of the solar neutrinos and they
reported that the LMA MSW solution
gives the best fit to the data \cite{ysuzuki}.
At the same time they also showed that the scenario of
pure sterile neutrino oscillations
$\nu_e\leftrightarrow\nu_s$ is excluded at 95\%CL.
It has been known that the atmospheric neutrino anomaly can be
accounted for by dominant $\nu_\mu\leftrightarrow\nu_\tau$
oscillations with almost maximal mixing
$(\Delta m_{\mbox{\rm\footnotesize atm}}^2,
\sin^22\theta_{\mbox{\rm\footnotesize atm}})\simeq (10^{-2.5}{\rm
eV}^2,1.0)$.  Again the Superkamiokande group has announced \cite{SKs}
that the possibility of pure sterile neutrino oscillations
$\nu_\mu\leftrightarrow\nu_s$ is excluded at 99\%CL.
On the other hand, combining the final result of LSND and the negative
results by E776 \cite{e776} ($\nu_\mu\rightarrow\nu_e$),
Karmen2 \cite{karmen} ($\nu_\mu\rightarrow\nu_e$) and Bugey \cite{bugey}
(${\bar\nu}_e\rightarrow{\bar\nu_e}$),
the oscillation parameter satisfies
0.1 eV$^2\lesssim\Delta m^2\lesssim 8$ eV$^2$ and
$8\times 10^{-4}\lesssim\sin^22\theta\lesssim 0.04$ at 99\%CL.
In this talk I will review the present status of four neutrino
scenarios in the light of the recent Superkamiokande data
of the solar and atmospheric neutrinos.

\section{Mass patterns}
\begin{figure}
\begin{center}
\vglue -0.5cm \hglue -0.5cm
\epsfig{file=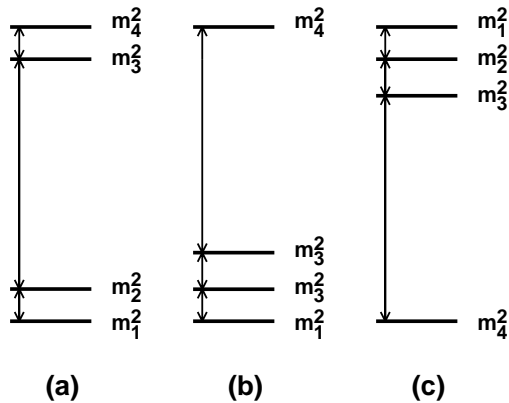,width=6cm}
\vglue 0.5cm
\caption{Mass patterns of four neutrino schemes.
(a) corresponds to (2+2)-scheme, where either
($|\Delta m^2_{21}|=\Delta m^2_\odot$,
$|\Delta m^2_{43}|=\Delta m^2_{\mbox{\rm\footnotesize atm}}$)
or
($|\Delta m^2_{43}|=\Delta m^2_\odot$,
$|\Delta m^2_{21}|=\Delta m^2_{\mbox{\rm\footnotesize atm}}$).
(b) and (c) are (3+1)-scheme, where
$|\Delta m^2_{41}|=\Delta m^2_{\mbox{\rm{\scriptsize LSND}}}$ and
either
($|\Delta m^2_{21}|=\Delta m^2_\odot$,
$|\Delta m^2_{32}|=\Delta m^2_{\mbox{\rm\footnotesize atm}}$)
or
($|\Delta m^2_{32}|=\Delta m^2_\odot$,
$|\Delta m^2_{21}|=\Delta m^2_{\mbox{\rm\footnotesize atm}}$) is satisfied.}
\label{fig:pattern}
\end{center}
\end{figure}

In the case of four neutrino schemes there are two distinct types of
mass patterns.  One is the so-called (2+2)-scheme
(Fig. \ref{fig:pattern}(a)) and the other is the (3+1)-scheme
(Fig. \ref{fig:pattern}(b) or (c)).  Depending on the type of the two
schemes, phenomenology is different.

\subsection{(3+1)-scheme}

It has been shown in Refs.~\cite{oy,bgg} using older data of LSND
\cite{lsnd} that the (3+1)-scheme is inconsistent with the Bugey
reactor data\cite{bugey} and the CDHSW disappearance experiment\cite{cdhsw}
of $\nu_\mu$.  Let me briefly give this argument in Refs.~\cite{oy,bgg}.
Without loss of generality I assume that one distinct mass eigenstate
is $\nu_4$ (See Fig. \ref{fig:pattern}(b) or (c)) and the largest mass
squared difference is
$\Delta m^2_{43}\equiv\Delta m^2_{\mbox{\rm{\scriptsize LSND}}}$.

\begin{figure}
\begin{center}
\vglue 5.0cm \hglue 1.2cm
\epsfig{file=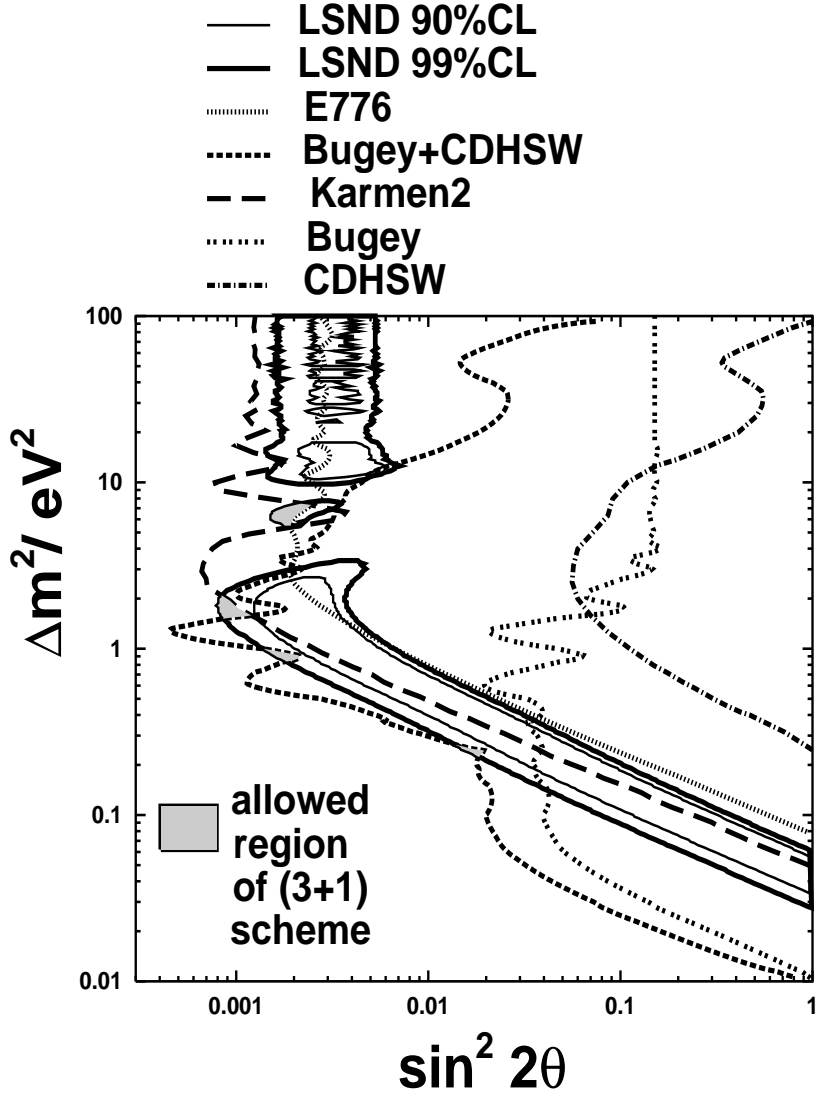,width=9cm}
\vglue -0.8cm
\caption{The allowed region of LSND from the final result
(the inside of the thick and thin solid lines corresponds to
the allowed region at 99\%CL and 90\%CL, respectively)
and the excluded regions of E776, Karmen2, Bugey, CDHSW
(the right hand side of each line).
The right hand side of the line (Bugey+CDHSW) stands for the excluded region
in the case of the (3+1)-scheme.  Only the four isolated shadowed areas
at $\Delta m^2_{\mbox{\rm{\scriptsize LSND}}}\simeq$0.3, 0.9, 1.7, 6.0 eV$^2$
are consistent with the LSND allowed region at 99\%CL in the (3+1)-scheme.}
\label{fig:lsnd}
\end{center}
\end{figure}

In the case of (3+1)-scheme the constraints from the Bugey and CDHSW
data are given by
\begin{eqnarray}
1-P({\bar\nu}_e\rightarrow{\bar\nu}_e)&=&
4|U_{e4}|^2(1-|U_{e4}|^2)\Delta_{43}
\le\sin^22\theta_{\mbox{\rm\scriptsize Bugey}}(\Delta m^2_{43})
\Delta_{43},\nonumber\\
1-P(\nu_\mu\rightarrow\nu_\mu)&=&
4|U_{\mu4}|^2(1-|U_{\mu4}|^2)\Delta_{43}
\le\sin^22\theta_{\mbox{\mbox{\rm{\scriptsize CDHSW}}}}
(\Delta m^2_{43})\Delta_{43},\nonumber
\end{eqnarray}
respectively, where $\Delta_{43}\equiv\sin^2
\left(\Delta m^2_{43}L/4E\right)$,
$\sin^22\theta_{\mbox{\rm\scriptsize Bugey}}$ and
$\sin^22\theta_{\mbox{\mbox{\rm{\scriptsize CDHSW}}}}$
stand for the values of the boundary
of the excluded region in the two flavor analysis as functions
of $\Delta m^2$ (See Fig. \ref{fig:lsnd}).
To explain the solar neutrino deficit and the zenith angle dependence
of the atmospheric neutrino data it is necessary to have
$|U_{e4}|^2<1/2$ and $|U_{\mu4}|^2<1/2$ and therefore I get
\begin{eqnarray}
|U_{e4}|^2&\le&{1 \over 2}
\left[1-\sqrt{1-\sin^22\theta_{\mbox{\rm\scriptsize Bugey}}
(\Delta m^2_{43})}\right]
\nonumber\\
|U_{\mu4}|^2&\le&{1 \over 2}
\left[1-\sqrt{1-\sin^22\theta_{\mbox{\mbox{\rm{\scriptsize CDHSW}}}}
(\Delta m^2_{43})}\right].
\label{eqn:emu1}
\end{eqnarray}
On the other hand, the appearance probability
$P(\bar{\nu}_\mu\rightarrow\bar{\nu}_e)$ of LSND in our scenario
is given by
\begin{eqnarray}
P(\bar{\nu}_\mu\rightarrow\bar{\nu}_e)&=&
4|U_{e4}|^2|U_{\mu4}|^2\Delta_{43}
\equiv\sin^22\theta_{\mbox{\rm{\scriptsize LSND}}}(\Delta m^2_{43})
\Delta_{43},
\label{eqn:emu2}
\end{eqnarray}
where $\sin^22\theta_{\mbox{\rm{\scriptsize LSND}}}(\Delta m^2_{43})$
stands for
the value of $\sin^22\theta$ in the LSND allowed region
in the two flavor framework.  From (\ref{eqn:emu1}) and
(\ref{eqn:emu2}) I obtain
\begin{eqnarray}
\sin^22\theta_{\mbox{\rm{\scriptsize LSND}}}(\Delta m^2_{43})&\le&
\left[1-\sqrt{1-\sin^22\theta_{\mbox{\rm\scriptsize Bugey}}
(\Delta m^2_{43})}\right]
\nonumber\\
&\times&
\left[1-\sqrt{1-\sin^22\theta_{\mbox{\mbox{\rm{\scriptsize CDHSW}}}}
(\Delta m^2_{43})}\right].
\label{eqn:emu3}
\end{eqnarray}
The value of the right hand side of (\ref{eqn:emu3}) is plotted in
Fig. \ref{fig:lsnd} together with the allowed region of LSND \cite{mills}.
At 90\%CL the allowed region of LSND does not satisfy the condition
(\ref{eqn:emu3}) for the (3+1)-scheme, and actually it used to be the
case with older data of LSND \cite{lsnd} even at 99\%CL \cite{oy,bgg}.
However, in the final result the allowed region has shifted
to the lower value of $\sin^22\theta$ and it was shown \cite{bklw}
that there are four isolated regions
$\Delta m^2_{\mbox{\rm{\scriptsize LSND}}}\simeq$0.3, 0.9, 1.7, 6.0 eV$^2$
which satisfy the condition
(\ref{eqn:emu3}).

\begin{figure}
\vglue -0.5cm
\hglue 4.7cm \epsfig{file=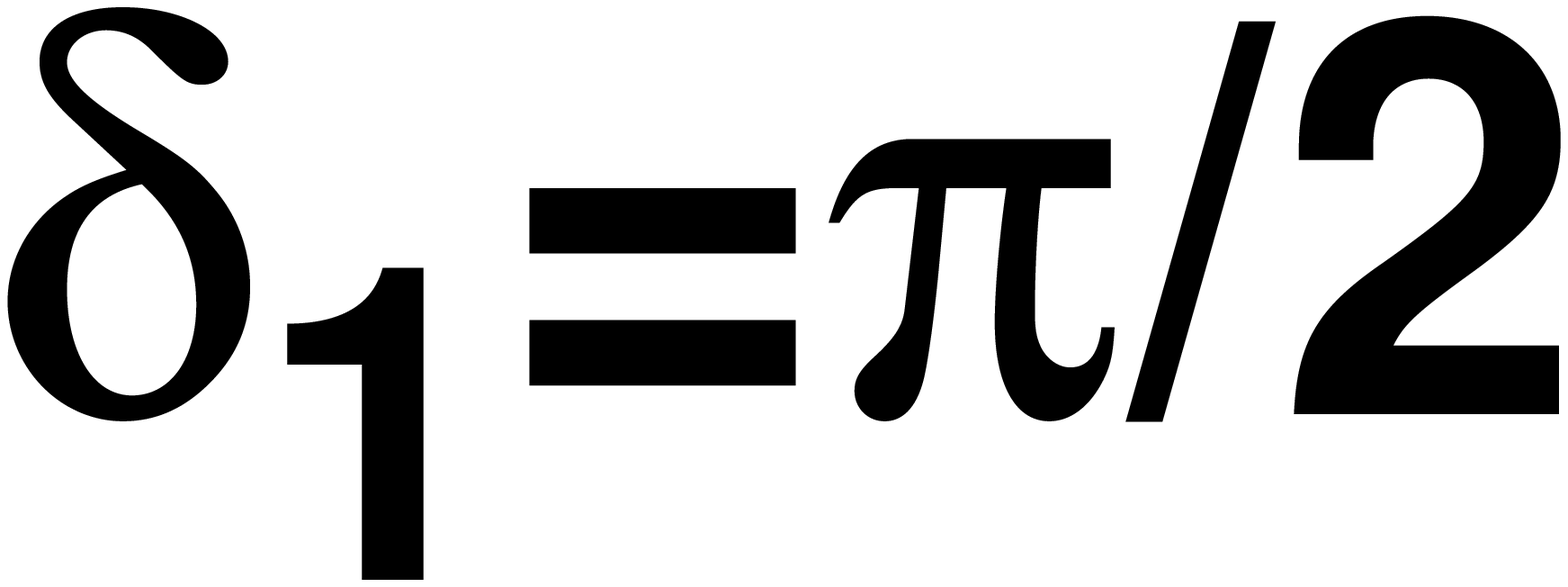,width=1.5cm}
\vglue -1.5cm
\hglue -2.0cm 
\epsfig{file=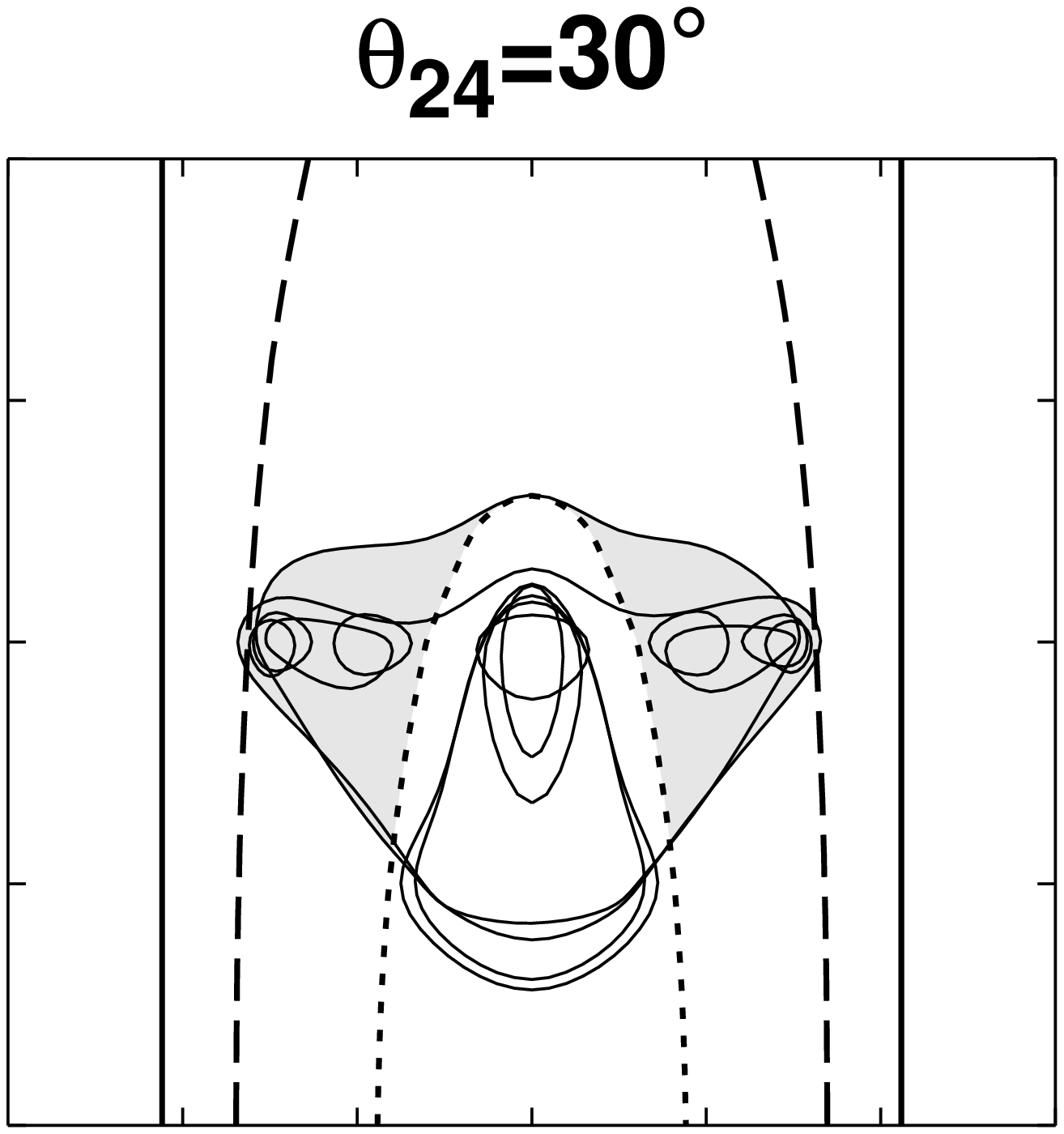,width=7cm}
\vglue -7.05cm \hglue 2.5cm \epsfig{file=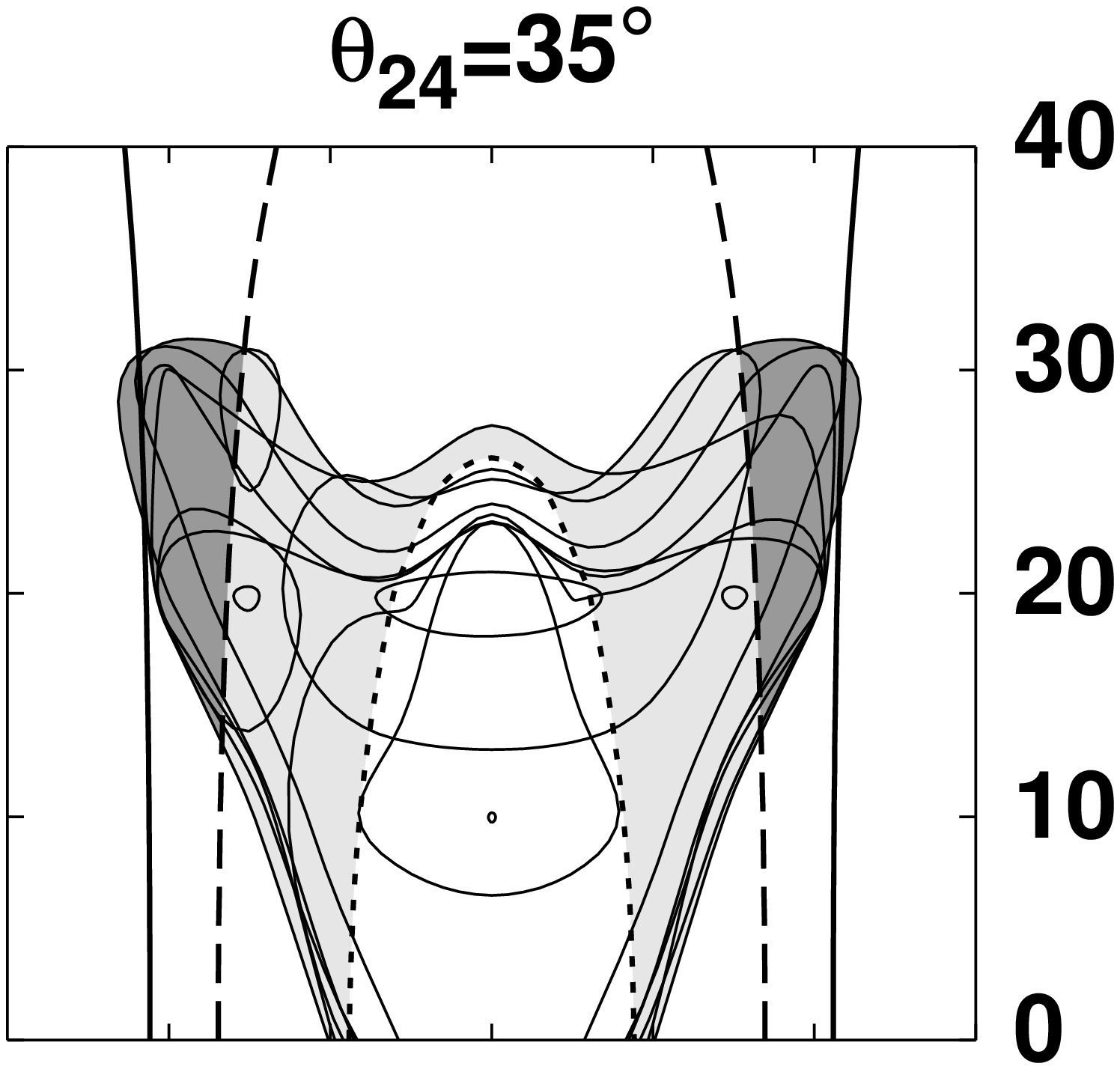,width=7cm}

\vglue -2.8cm
\hglue -2.0cm 
\epsfig{file=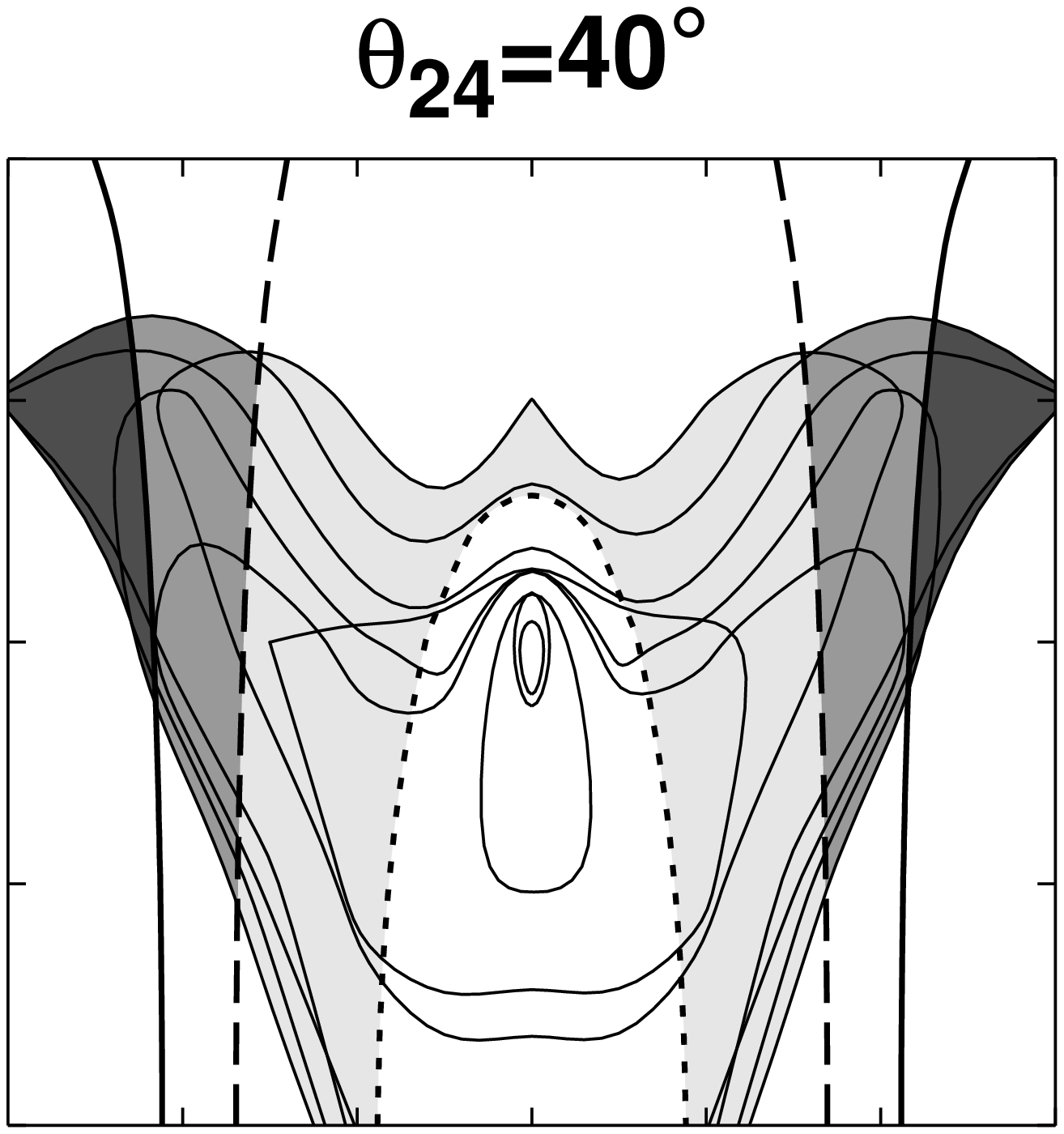,width=7cm}
\vglue -7cm \hglue 2.5cm \epsfig{file=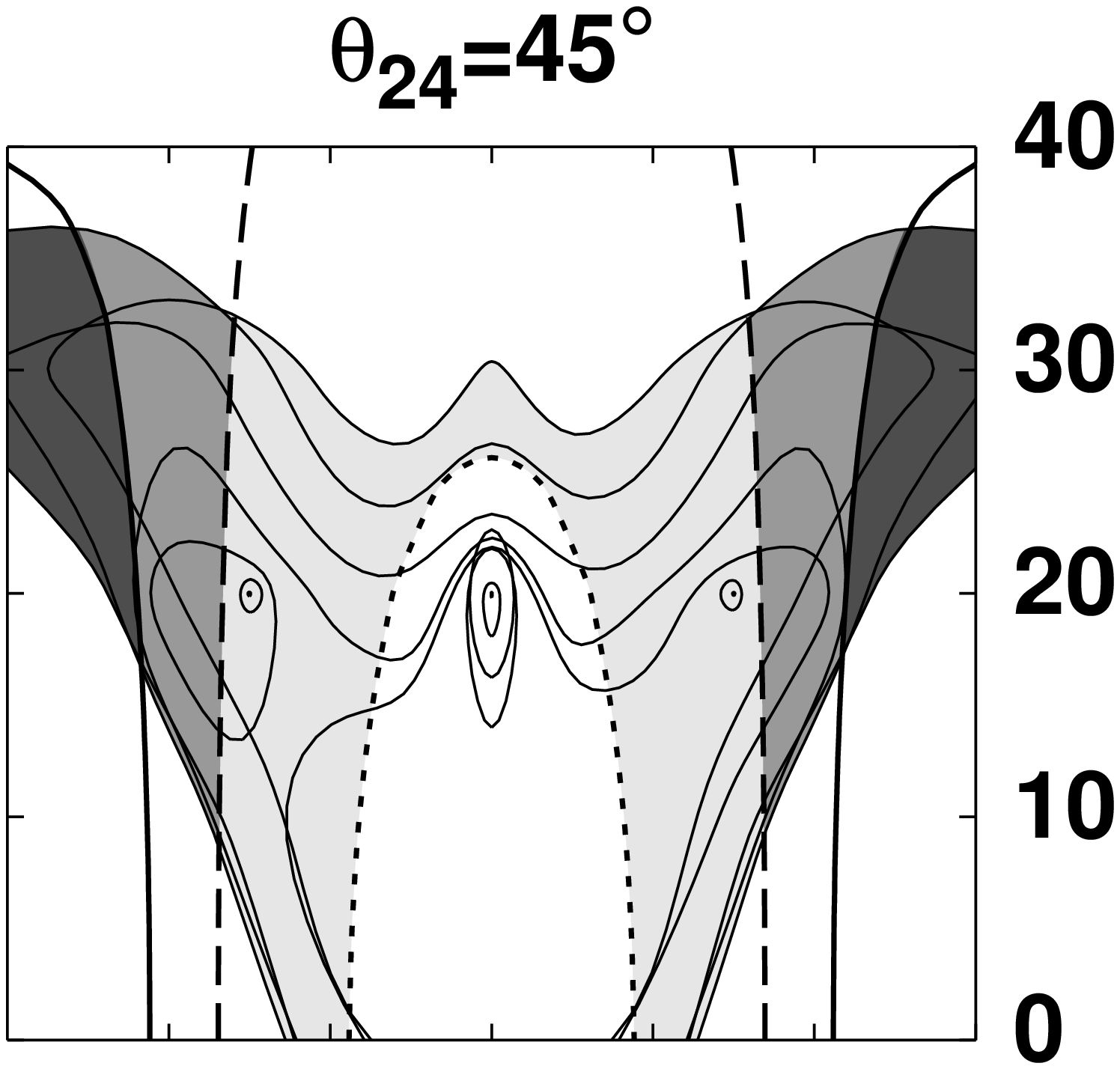,width=7cm}

\vglue -2.8cm
\hglue -2.0cm 
\epsfig{file=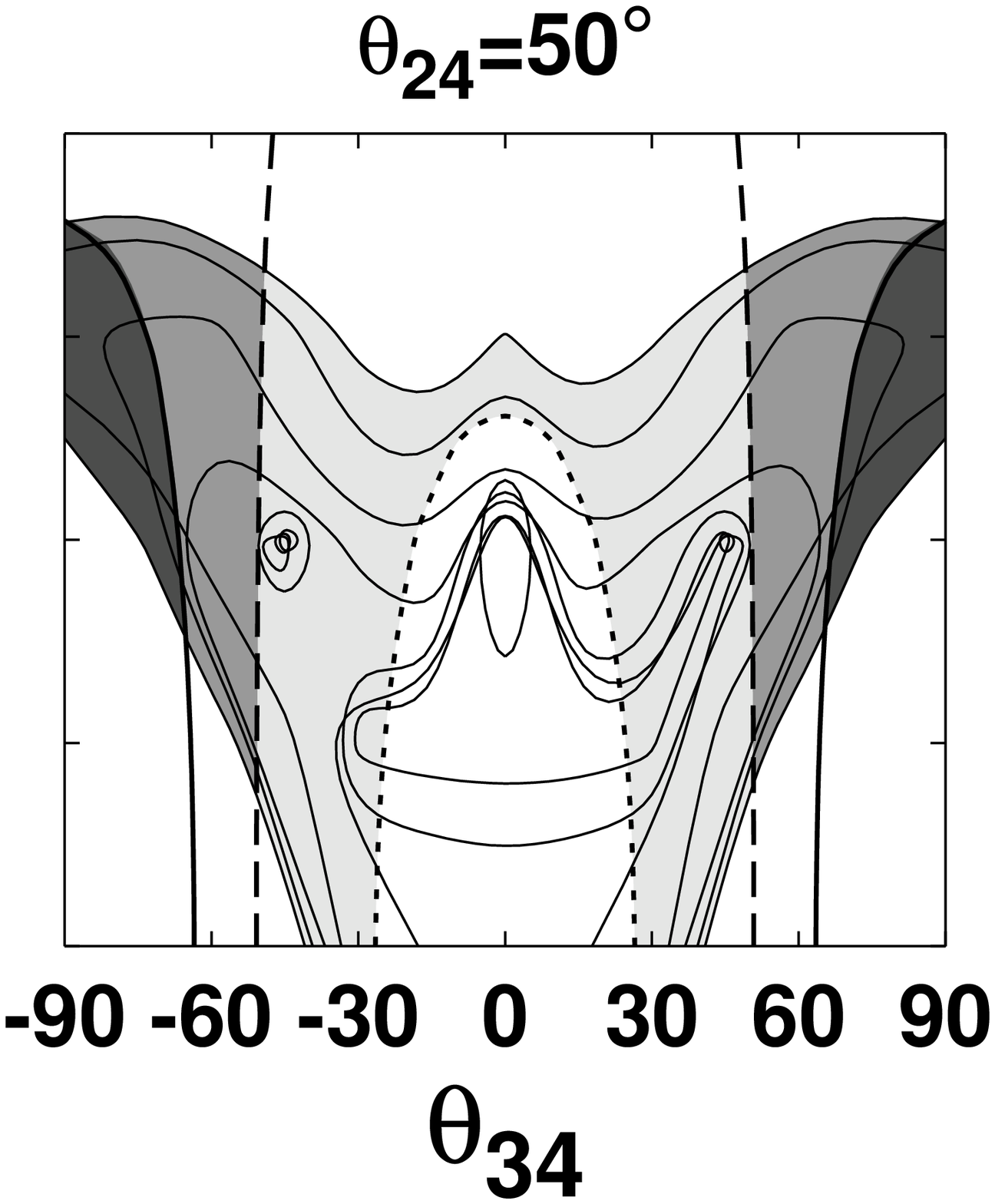,width=7cm}
\vglue -7cm \hglue 2.5cm \epsfig{file=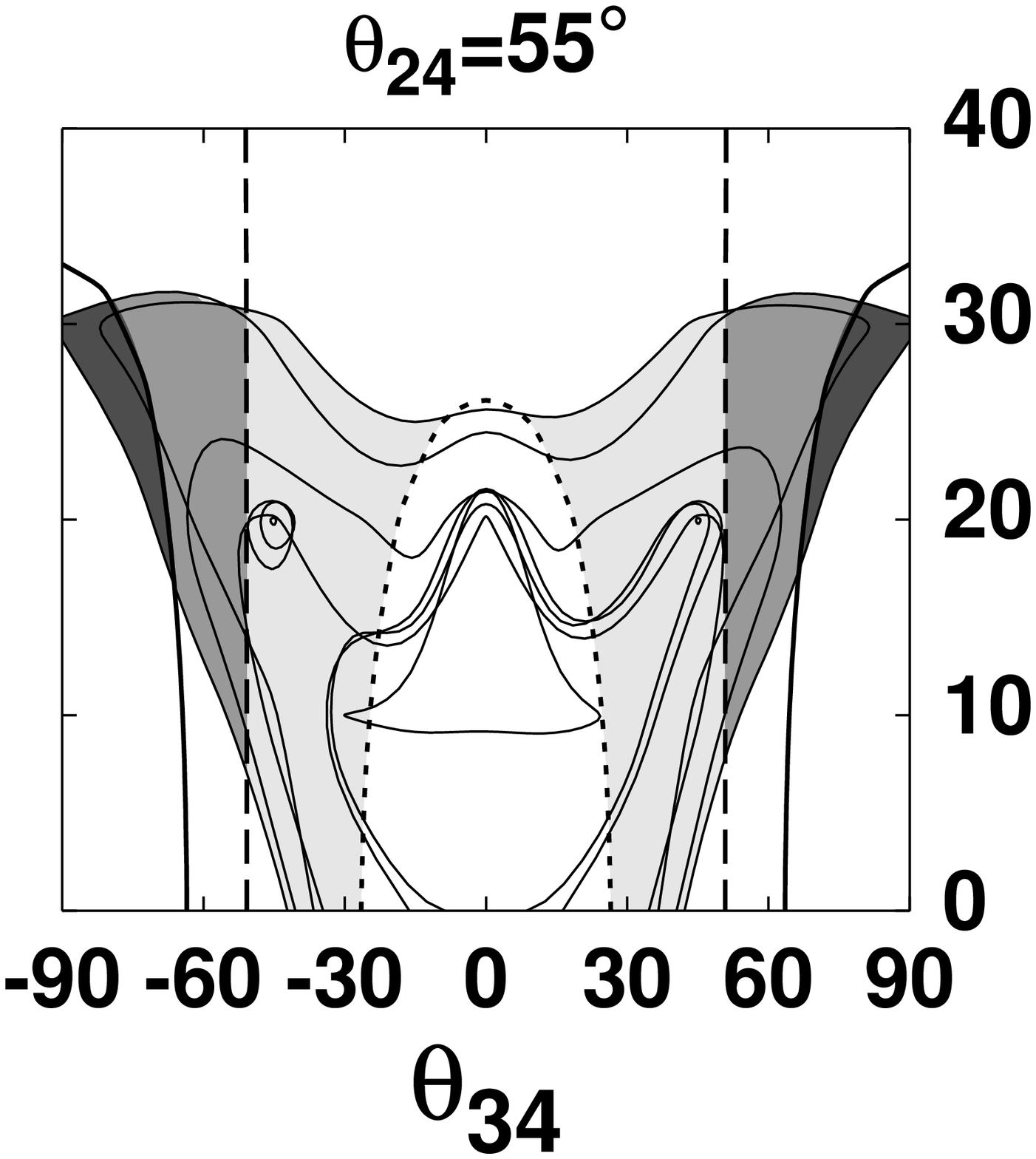,width=7cm}
\vglue -1.0cm \hglue 1.5cm \epsfig{file=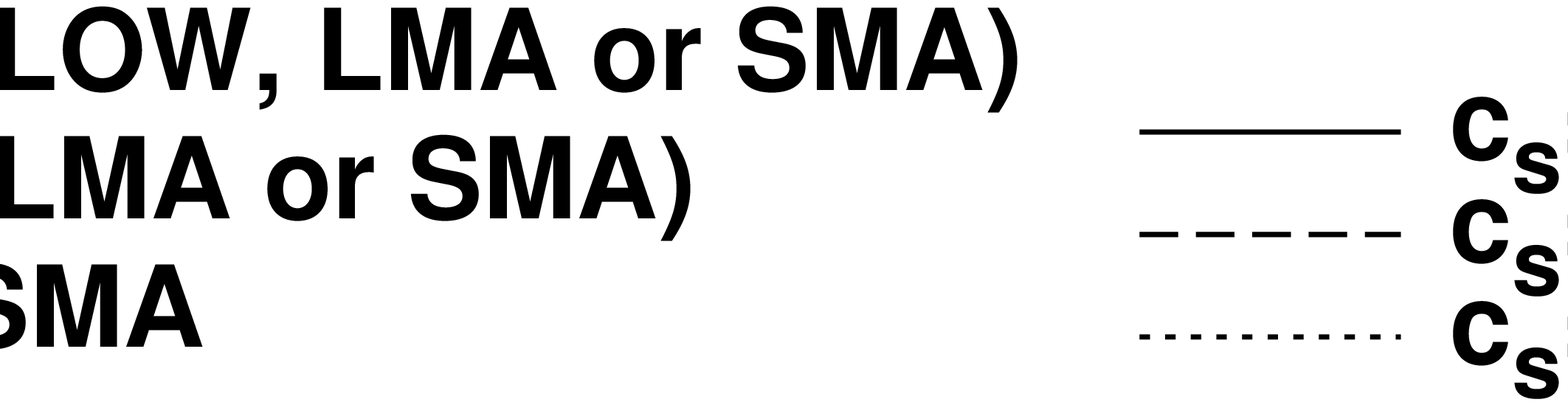,width=7cm}
\vglue -5.0cm
\caption{The allowed region of the (2+2)-scheme with
$\delta_1=\pi/2$ at 90 \%CL constrained
by the Superkamiokande atmospheric neutrino data.
The shadowed regions stand for the allowed regions from
the combined analysis of the solar and atmospheric neutrino
data.}
\label{fig:atm4}
\end{figure}

\subsection{(2+2)-scheme}

In the case of the (2+2)-scheme, assuming the mass pattern in
Fig. \ref{fig:pattern} (a) with
$\Delta m^2_{21}=\Delta m^2_\odot$,
$\Delta m^2_{32}=\Delta m^2_{\mbox{\rm\footnotesize atm}}$,
$\Delta m^2_{43}=\Delta m^2_{\mbox{\rm{\scriptsize LSND}}}$,
the constraints from the LSND, Bugey and CDHSW
data are given by
\begin{eqnarray}
P(\bar{\nu}_\mu\rightarrow\bar{\nu}_e)&=&
4|U_{e3}U^\ast_{\mu3}+U_{e4}U^\ast_{\mu4}|^2\Delta_{32}
\equiv\sin^22\theta_{\mbox{\rm{\scriptsize LSND}}}(\Delta m^2_{32})
\Delta_{32},\nonumber\\
1-P({\bar\nu}_e\rightarrow{\bar\nu}_e)&=&
4(|U_{e3}|^2+|U_{e4}|^2)(1-|U_{e3}|^2-|U_{e4}|^2)
\Delta_{32}\nonumber\\
&\le&\sin^22\theta_{\mbox{\rm\scriptsize Bugey}}(\Delta m^2_{32})
\Delta_{32},\nonumber\\
1-P(\nu_\mu\rightarrow\nu_\mu)&=&
4(|U_{\mu3}|^2+|U_{\mu4}|^2)(1-|U_{\mu3}|^2-|U_{\mu4}|^2)
\Delta_{32}\nonumber\\
&\le&\sin^22\theta_{\mbox{\mbox{\rm{\scriptsize CDHSW}}}}
(\Delta m^2_{32})\Delta_{32},
\label{eqn:cdhsw}
\end{eqnarray}
where $\Delta_{32}\equiv\sin^2
\left(\Delta m^2_{32}L/4E\right)$.

It has been shown \cite{oy,bgg} that these conditions are consistent
with all the constraints from the accelerator, reactor data as well as
solar and atmospheric neutrino observations.  As I will show, to
account for both the solar neutrino deficit and the atmospheric
neutrino anomaly, it is necessary to have
\begin{eqnarray}
4(|U_{\mu3}|^2+|U_{\mu4}|^2)(1-|U_{\mu3}|^2-|U_{\mu4}|^2)\sim
{\cal O}(1),
\label{eqn:cond1}
\end{eqnarray}
so I take
$\Delta m^2_{32}$ as small as possible, i.e., $\Delta m^2_{32}=0.3$ eV$^2$
so that (\ref{eqn:cond1}) be consistent with the CDHSW constraint
(\ref{eqn:cdhsw}).

\section{(2+2)-scheme}
\subsection{Analysis of the solar neutrino data}
The solar neutrino data were analyzed in the framework
of the (2+2)-scheme by Ref.~\cite{ggp} on the assumption that
$U_{e3}=U_{e4}=0$, which is justified from the Bugey
constraint $|U_{e3}|^2+|U_{e4}|^2\ll 1$, and
$\Delta m^2_{\mbox{\rm\footnotesize atm}}$,
$\Delta m^2_{\mbox{\rm\scriptsize LSND}}\rightarrow\infty$ which is also
justified since $|\Delta m^2_{\mbox{\rm\footnotesize atm}}/2E|$,
$|\Delta m^2_{\mbox{\rm\scriptsize LSND}}/2E|\gg\sqrt{2}G_FN_e$
for the solar neutrino problem, where $G_F$ and $N_e$
stand for the Fermi constant and the electron density in the Sun.
The conclusion of Ref.~\cite{ggp}is that
the SMA MSW solution exists for
$0\le c_s \lesssim 0.8$, while the
LMA MSW and LOW solutions survive only for $0\le c_s
\lesssim 0.4$ and $0\le c_s \lesssim 0.2$, respectively,
where $c_s\equiv|U_{s1}|^2+|U_{s2}|^2$.

\subsection{Analysis of the atmospheric neutrino data}

The atmospheric neutrino data were analyzed by Refs.~\cite{y1,flm}
with the (2+2)-scheme.  Refs.~\cite{y1,flm} assumed $U_{e3}=U_{e4}=0$
as in Ref.~\cite{ggp}, and $\Delta m^2_\odot=0$ was also assumed.
Ref.~\cite{y1} assumed $\Delta m^2_{\mbox{\rm{\scriptsize LSND}}}=0.3$ eV$^2$
so that
the result with large $|U_{\mu3}|^2+|U_{\mu4}|^2$ do not contradict
with the CDHSW constraint (\ref{eqn:cdhsw}).  Ref.~\cite{flm} did not
take into account the contribution from
$\Delta m^2_{\mbox{\rm{\scriptsize LSND}}}$
to the oscillation probability and their result is a subset of
Ref.~\cite{y1}.

Here I adopt the notation in Ref.~\cite{oy} for the $4\times 4$ MNS matrix:
\begin{eqnarray}
U_{MNS}&\equiv& R_{34}({\pi \over 2}-\theta_{34})R_{24}(\theta_{24})
R_{23}({\pi \over 2})
U_{23}(\theta_{23},\delta_1)
U_{14}(\theta_{14},\delta_3)
U_{13}(\theta_{13},\delta_2)
R_{12}(\theta_{12})\nonumber\\
\label{eqn:mns}
\end{eqnarray}
where
$U_{23}(\theta_{23},\delta_1)\equiv
e^{2i\delta_1\lambda_3}R_{23}(-\theta_{23})
e^{-2i\delta_1\lambda_3}$,
$U_{14}(\theta_{14},\delta_3)\equiv
e^{\sqrt{6}i\delta_3\lambda_{15}/2}
R_{14}(\theta_{14})$
$e^{-\sqrt{6}i\delta_3\lambda_{15}/2}$,
$U_{13}(\theta_{13},\delta_2)\equiv
e^{2i\delta_2\lambda_8/\sqrt{3}}R_{13}(\theta_{13})
e^{-2i\delta_2\lambda_8/\sqrt{3}}$,
$R_{jk}(\theta)\equiv \exp\left(iT_{jk}\theta\right),$
$\left(T_{jk}\right)_{\ell m}=i\left(\delta_{j\ell}\delta_{km}
-\delta_{jm}\delta_{k\ell}\right)$,
$2\lambda_3\equiv{\rm diag}(1,-1,0,0)$,
$2\sqrt{3}\lambda_8\equiv{\rm diag}(1,1,-2,0)$,
$2\sqrt{6}\lambda_{15}\equiv{\rm diag}(1,1,1,-3)$
are $4\times4$ matrices
($\lambda_j$ are elements of the $su(4)$ generators).
With the assumptions $\Delta m^2_{21}=0$,
$|U_{e4}|^2=s^2_{14}=0$, $|U_{e3}|^2=c^2_{14}s^2_{13}=0$,
$\theta_{12}$, $\theta_{13}$ and $\theta_{14}$ disappear
from $U$ and $\nu_e$ decouples from other three neutrinos.
Thus the problem is reduced to the three flavor neutrino
analysis among $\nu_\mu$, $\nu_\tau$, $\nu_s$ and the reduced
MNS matrix is
\begin{eqnarray}
\widetilde U&\equiv&\left(
\begin{array}{ccc}
 U_{\mu 2} & U_{\mu 3}&U_{\mu 4} \\
 U_{\tau 2} & U_{\tau 3}&U_{\tau 4} \\
 U_{s2} &  U_{s3}&U_{s4} 
\end{array}\right)
=e^{i({\pi \over 2}-\theta_{34})\lambda_7}
D^{-1} e^{i\theta_{24}\lambda_5} D~
e^{i(\theta_{23}-{\pi \over 2})\lambda_2},\nonumber
\end{eqnarray}
with $D\equiv{\rm diag}\left(e^{i\delta_1/2},1,e^{-i\delta_1/2} \right)$
($\lambda_j$ are the $3\times 3$ Gell-Mann matrices)
is the reduced $3\times 3$ MNS matrix.
This MNS matrix $\widetilde U$
is obtained by substitution
$\theta_{12}\rightarrow\theta_{23}-\pi/2$,
$\theta_{13}\rightarrow\theta_{24}$,
$\theta_{12}\rightarrow\pi/2-\theta_{34}$,
$\delta\rightarrow\delta_1$
in the standard parametrization in Ref.~\cite{pdg}.
It turns out that $\theta_{34}$ corresponds to the
mixing of $\nu_\mu\leftrightarrow\nu_\tau$ and
$\nu_\mu\leftrightarrow\nu_s$, while $\theta_{23}$
is the mixing of the contribution of
$\displaystyle\sin^2\left(\Delta m^2_{\mbox{\rm\footnotesize atm}}
L/4E\right)$
and
$\displaystyle\sin^2\left(\Delta m^2_{\mbox{\rm{\scriptsize LSND}}}
L/4E\right)$
in the oscillation probability.  The allowed region at 90\%CL
of the atmospheric neutrino data is given by the area bounded
by thin solid lines in Fig. \ref{fig:atm4} for $\delta_1=\pi/2$.
The allowed regions for $\delta_1=0, \pi/4$ are given in Ref.~\cite{y1}.

\subsection{Combined analysis of the solar and atmospheric neutrino data}
In Fig. \ref{fig:atm4}, the lines given by
$c_s\equiv|U_{s1}|^2+|U_{s2}|^2=|c_{23}c_{34}
+s_{23}s_{34}s_{24}e^{i\delta_1}|^2=0.2, 0.4, 0.8$ are depicted
together with the allowed region of the atmospheric neutrino data.
By combining the analyses of Ref.~\cite{ggp} and Ref.~\cite{y1},
I obtain the region which satisfies the constraints of the
solar and atmospheric neutrino data.
The darkest, medium and lightest shadowed areas stand for 
$\nu_{\mbox{\rm\footnotesize atm}}+\nu_\odot$(SMA, LMA or LOW),
$\nu_{\mbox{\rm\footnotesize atm}}+\nu_\odot$(SMA or LMA),
$\nu_{\mbox{\rm\footnotesize atm}}+\nu_\odot$(SMA),
respectively.
Although this result is not quantitative, it gives us a sense
on how likely the (2+2)-scheme is allowed by combining the solar
and atmospheric neutrino data.
Let me emphasize that non-zero contribution of
$\displaystyle\sin^2\left(\Delta m^2_{\mbox{\rm{\scriptsize LSND}}}
L/4E\right)$
(i.e., the case of $\theta_{23}>0$) to the oscillation probability
is important particularly for the LMA and LOW solar solutions.
The region of $\theta_{23}>0$ has not been
analyzed by Ref.~\cite{flm}.  Let me also stress that
both the solar neutrinos and the atmospheric neutrinos
are accounted for by hybrid of active and sterile oscillations
in the (2+2)-scheme.

\section{(3+1)-scheme}
After the work of Barger et al. \cite{bklw}, people \cite{gl,ps} have
investigated various consequences of the (3+1)-scheme.
Here let me make two comments on the (3+1)-scheme.

\subsection{Atmospheric neutrinos}
As in the case of the (2+2)-scheme, I assume
$U_{e3}=U_{e4}=\Delta m^2_\odot=0$ for simplicity.
Then $\nu_e$ once again decouples from $\nu_e$, $\nu_\mu$, $\nu_\tau$ 
and the probability in vacuum
for the atmospheric neutrino scale
is given by
\begin{eqnarray}
P(\nu_\mu\rightarrow\nu_\mu)&=&
4|U_{\mu3}|^2(1-|U_{\mu3}|^2-|U_{\mu4}|^2)\Delta_{32}
+2|U_{\mu4}|^2(1-|U_{\mu4}|^2),\nonumber\\
P(\nu_\mu\rightarrow\nu_\tau)&=&
4\Re\left[U_{\mu3}U^\ast_{\tau3}(U^\ast_{\mu3}U_{\tau3}
+U^\ast_{\mu4}U_{\tau4})\right]\Delta_{32}
+2|U_{\mu4}|^2|U_{\tau4}|^2,\nonumber\\
P(\nu_\mu\rightarrow\nu_s)&=&
4\Re\left[U_{\mu3}U^\ast_{s3}(U^\ast_{\mu3}U_{s3}
+U^\ast_{\mu4}U_{s4})\right]\Delta_{32}
+2|U_{\mu4}|^2|U_{s4}|^2,\nonumber\\
\end{eqnarray}
where I have taken
$\Delta m^2_{32}\equiv\Delta m^2_{\mbox{\rm\footnotesize atm}}$,
$\Delta m^2_{43}\equiv\Delta m^2_{\mbox{\rm{\scriptsize LSND}}}$
and I have averaged
over rapid oscillations: $\sin^2
\left({\Delta m^2_{\mbox{\rm{\scriptsize LSND}}}L/4E}\right)\rightarrow1/2$.
Since the (3+1)-scheme is allowed only for four discrete
values of $\Delta m^2_{\mbox{\rm{\scriptsize LSND}}}$, let me discuss
$\Delta m^2_{\mbox{\rm{\scriptsize LSND}}}$=0.3 eV$^2$
($|U_{\mu4}|^2\gtrsim0.34$)
and $\Delta m^2_{\mbox{\rm{\scriptsize LSND}}}$=0.9 eV$^2$ 
($|U_{\mu4}|^2\simeq0.03$),
1.7 eV$^2$ ($|U_{\mu4}|^2\simeq0.01$), 6.0 eV$^2$ ($|U_{\mu4}|^2\simeq0.02$),
separately.  For simplicity I assume $\delta_1$=0 since the existence
of the CP phase $\delta_1$ does not change the situation very much.

\subsubsection{$\Delta m^2_{\mbox{\rm{\scriptsize LSND}}}$=0.3 eV$^2$}

Since we know from the Superkamiokande atmospheric neutrino
data that the coefficient of $\sin^2
\left({\Delta m^2_{\mbox{\rm\footnotesize atm}}L/4E}\right)$ in
$P(\nu_\mu\rightarrow\nu_\mu)$ has to be large to have a good fit,
I optimize $4|U_{\mu3}|^2(1-|U_{\mu3}|^2-|U_{\mu4}|^2)$
with respect to $\theta_{23}$ for $|U_{\mu4}|^2=0.34$.
When $U_{e4}=0$ I have $|U_{\mu3}|^2=c^2_{23}c^2_{24}$,
$|U_{\mu4}|^2=s^2_{24}$,
$|U_{\tau4}|^2=c^2_{24}c^2_{34}$, $|U_{s4}|^2=c^2_{24}s^2_{34}$
in the notation of Ref.~\cite{oy}, and
it is easy to see
\begin{eqnarray}
4|U_{\mu3}|^2(1-|U_{\mu3}|^2-|U_{\mu4}|^2)
=c^4_{24}\sin^22\theta_{23}\le c^4_{24}=0.44,\nonumber
\end{eqnarray}
where equality holds when $\theta_{23}=\pi/4$.  This is the value
of $\theta_{23}$ for which the fit of the (3+1)-scheme
to the atmospheric neutrino data is supposed
to be the best for $\Delta m^2_{\mbox{\rm{\scriptsize LSND}}}$=0.3 eV$^2$.
When $\theta_{23}=\pi/4$
the probability in vacuum becomes
\begin{eqnarray}
P(\nu_\mu\rightarrow\nu_\tau)&=&
\left(c^2_{24}s^2_{34}-{1 \over 4}\sin^22\theta_{24}c^2_{34}\right)
\Delta_{32}
+{1 \over 2}c^2_{34}\sin^22\theta_{24}\nonumber\\
P(\nu_\mu\rightarrow\nu_s)&=&
\left(c^2_{24}c^2_{34}-{1 \over 4}\sin^22\theta_{24}s^2_{34}\right)
\Delta_{32}
+{1 \over 2}s^2_{34}\sin^22\theta_{24}.\nonumber\\
\label{eqn:pmutau}
\end{eqnarray}
From (\ref{eqn:pmutau}) $\theta_{34}$ turns out to be
the mixing of $\nu_\mu\leftrightarrow\nu_\tau$ and
$\nu_\mu\leftrightarrow\nu_s$ as in the (2+2)-scheme.

I found from the explicit numerical calculation \cite{y2}
that the fit of the (3+1)-scheme with
$\Delta m^2_{\mbox{\rm{\scriptsize LSND}}}$=0.3 eV$^2$,
$|U_{\mu4}|^2=0.34$, $\theta_{23}=\pi/4$
to the atmospheric neutrino data is very bad for any
value of $\theta_{34}$ and
the region of $\Delta m^2_{\mbox{\rm{\scriptsize LSND}}}$=0.3 eV$^2$
is excluded at 6.9$\sigma$CL.

\subsubsection{$\Delta m^2_{\mbox{\rm{\scriptsize LSND}}}$
=0.9, 1.7, 6.0 eV$^2$}
In this case $|U_{\mu4}|^2\lesssim0.03$ and I can put
$U_{\mu4}=0$ as a good approximation.  Then the constant part
in the oscillation probability disappears and this case is
reduced to the analysis in the (2+2)-scheme with $\theta_{23}=0$.
The allowed region at 90\%CL is given roughly by
$-\pi/4\lesssim\theta_{34}\lesssim\pi/4$,
$0.8\lesssim\sin^22\theta_{24}\le1$,
where $\theta_{34}$ and $\theta_{24}$ stand for
the mixing of $\nu_\mu\leftrightarrow\nu_\tau$ and
$\nu_\mu\leftrightarrow\nu_s$ and the mixing of atmospheric
neutrino oscillations, respectively.

\subsection{Oscillations of high energy neutrinos in matter}

\begin{figure}[h]
\begin{center}
\vglue -4.5cm 
\hglue -2.0cm
\epsfig{file=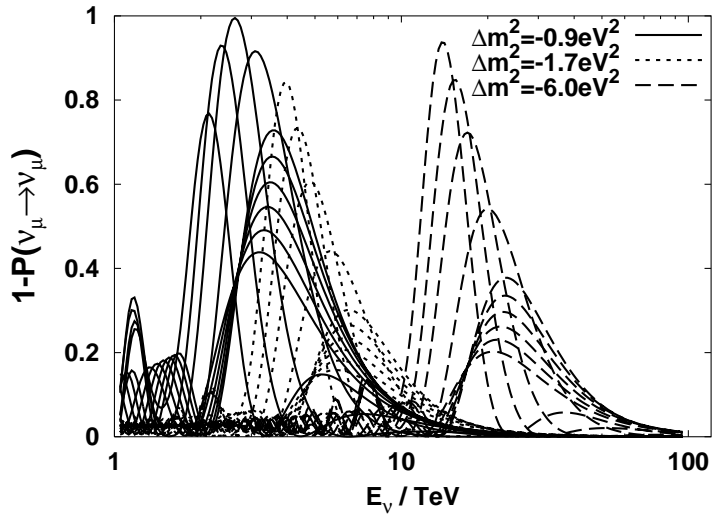,width=7cm}
\vglue 1.5cm 
\caption{Enhancement of neutrino oscillations due to matter effect in
the case of the (3+1)-scheme, where $\theta_{23}=45^\circ$ is taken
for all the cases.  For each value of $\Delta m^2$= -0.9 eV$^2$
($\epsilon=10.0^\circ$, $\delta=7.5^\circ$),
-1.7 eV$^2$ ($\epsilon=5.6^\circ$, $\delta=7.5^\circ$),
-6.0 eV$^2$ ($\epsilon=6.4^\circ$, $\delta=8.8^\circ$),
ten curves correspond to
$\cos\Theta=-1.0,-0.9,\cdots,-0.1$ from the left to the right, where the
zenith angle $\Theta$ is related to the baseline $L$ by $L=-2R\cos\Theta$ with
$R$=6378km.  Most of the channel is $\nu_\mu\rightarrow\nu_s$.}
\label{fig:p}
\end{center}
\end{figure}

When $|U_{e4}|^2$, $|U_{\mu4}|^2$ and $|U_{\tau4}|^2$ are
all small, it is naively difficult to distinguish
the (3+1)-scheme from the ordinary three flavor scenario.
However, because of the existence of the small mixing
angles in $U_{e4}$, $U_{\mu4}$ and the large mass squared
difference $\Delta m^2_{\mbox{\rm{\scriptsize LSND}}}$ the oscillation
probability in matter can have enhancement which never happens
in the three flavor case.  By taking $\theta_{12}=\pi/4$,
$\theta_{13}=0$, $\theta_{23}=\pi/4$,
$\theta_{14}=\epsilon$ ($|\epsilon|\ll 1$),
$\theta_{24}=\delta$ ($|\delta|\ll 1$),
$\theta_{34}=\pi/2$ in (\ref{eqn:mns}) I get
\begin{eqnarray}
U\simeq\left(
\begin{array}{cccc}
{1 \over \sqrt{2}} &  {\ }{\ }{1 \over \sqrt{2}}& {\ }0 & {\ }\epsilon\\
{1 \over 2}& {\ }-{1 \over \sqrt{2}} &  {\ }{1 \over \sqrt{2}} &{\ }\delta\\
-{1 \over 2} & {\ }{\ }{1 \over 2} &  {\ }{1 \over \sqrt{2}} & {\ }0\\
-{\epsilon \over \sqrt{2}}-{\delta \over 2}&
{\ }-{\epsilon \over \sqrt{2}}+{\delta \over 2} &   {\ }0 & {\ }1
\end{array}\right)
\left( \begin{array}{c} \nu_1  \\ \nu_2 \\ 
\nu_3\\\nu_4 \end{array} \right)\nonumber
\label{eqn:u}
\end{eqnarray}
which is the same as the MNS matrix in Ref.~\cite{bklw} up to the
phase of each factor.  The probability $P(\nu_\mu\rightarrow\nu_\mu)$
turns out to receive significant deviation from the
vacuum one due to the matter effect for $E_\nu\sim{\cal O}(1)$ TeV,
and the behaviors of $1-P(\nu_\mu\rightarrow\nu_\mu)$ are shown
in Fig. \ref{fig:p}, where three cases of
$|\Delta m^2_{\mbox{\rm{\scriptsize LSND}}}|$=0.9 eV$^2$
($8.8^\circ\le\epsilon\le12.2^\circ$,
$6.4^\circ\le\delta\le8.9^\circ$), 1.7 eV$^2$
($7.5^\circ\le\epsilon\le10.2^\circ$,
$5.6^\circ\le\delta\le7.7^\circ$), 6.0 eV$^2$
($7.5^\circ\le\epsilon\le7.7^\circ$,
$10.0^\circ\le\delta\le10.2^\circ$) are considered.\footnote{
The eigenvalues of the mass matrix in this case turn out to be
roots of a cubic equation and analytic treatment of the
oscillation probability is difficult, unlike the cases
of three flavors \cite{y3} or four flavors \cite{dgkk,gl},
where one mass scale is dominant
and the eigenvalues are roots of
a quadratic equation.}
The appearance channel which is enhanced is dominantly
$\nu_\mu\rightarrow\nu_s$, so it may be difficult to detect
signs of this enhancement from observations of high energy
cosmic neutrinos of energy $E_\nu\sim{\cal O}(1)$ TeV,
although this enhancement may be observed through neutral current
interactions in the future.\footnote{Similar enhancement has
been discussed in a different context by Ref.~\cite{nth}.
I thank Athar Husain for bringing my attention to Ref.~\cite{nth}.}

\section{Big Bang Nucleosynthesis}
It has been shown \cite {bbn} in the two flavor framework that if
sterile neutrino have oscillations with active ones and
if $\Delta m^2\sin^42\theta\gtrsim 3\times10^{-3}$ eV$^2$ is satisfied
then sterile neutrinos would have been in thermal equilibrium
and the number $N_\nu$ of light neutrinos in Big Bang Nucleosynthesis
(BBN) would be 4.  This argument was generalized to the
four neutrino case \cite{oy,bggs} and by imposing all the constraints
from accelerators, reactors, solar neutrinos and atmospheric neutrinos
as well as the BBN constraint $N_\nu<4.0$ it was concluded that
the only consistent four neutrino scenario is the (2+2)-scheme with
the MNS mixing matrix
\begin{eqnarray}
U_{MNS}
=\left(
\begin{array}{cccc}
U_{e1} & U_{e2} &  U_{e3} &  U_{e4}\\
U_{\mu 1} & U_{\mu 2} & U_{\mu 3} & U_{\mu 4}\\
U_{\tau 1} & U_{\tau 2} & U_{\tau 3} & U_{\tau 4}\\
U_{s1} & U_{s2} &  U_{s3} &  U_{s4}
\end{array}\right)
\simeq\left(
\begin{array}{cccc}
c_\odot & s_\odot & \epsilon & \epsilon\\
\epsilon & \epsilon&1/\sqrt{2} & 1/\sqrt{2} \\
\epsilon & \epsilon&-1/\sqrt{2} & 1/\sqrt{2} \\
-s_\odot & c_\odot & \epsilon & \epsilon\\
\end{array}\right),
\label{eqn:ubbn}
\end{eqnarray}
where $c_\odot\equiv\cos\theta_\odot$, $s_\odot\equiv\sin\theta_\odot$
and $\theta_\odot$ stands for the mixing angle of the SMA MSW solar
solution.  In this case $c_s\equiv|U_{s1}|^2+|U_{s2}|^2\simeq 1$
and the solar neutrino deficit would be accounted for by sterile
neutrino oscillations $\nu_e\leftrightarrow\nu_s$ with the SMA MSW solution
while the atmospheric neutrino anomaly would be by
$\nu_\mu\leftrightarrow\nu_\tau$ oscillations.
This scenario is obviously inconsistent with
the recent solar neutrino data by the Superkamiokande group,
and the argument which has lead to (\ref{eqn:ubbn}) has to be given up.

Fortunately the upper bound on $N_\nu$ has become
less stringent now and $N_\nu=4.0$ seems to be allowed.
Furthermore, it has been shown recently \cite{emmmp} that 
the combined analysis of BBN and the recent data
by BOOMERanG \cite{boomerang} and MAXIMA-1 \cite{maxima}
of the Cosmic Microwave
Background prefers higher value of $N_\nu$: $4\le N_\nu\le 13$.
Therefore all the four neutrino schemes of type (2+2) and (3+1)
seemed to be consistent with the BBN constraint.

On the other hand, it has been pointed out \cite{fv} in the two flavor
framework that for a certain range of the oscillation parameters
neutrino oscillations themselves create asymmetry between $\nu$ and
$\bar\nu$ and this asymmetry prevents $\nu_s$ from oscillating into
active neutrinos.  Although this analysis has not been generalized to
the four neutrino cases, even if the upper bound of $N_\nu$ becomes
less than 4.0 in the future, it might be still possible to have four
neutrino schemes which are consistent with the BBN constraint as well
as the solar and atmospheric neutrino data due to possible asymmetry
in $\nu$ and $\bar\nu$.

\section{Conclusions}

In this talk I have shown that there are still four neutrino scenarios
((2+2)- as well as (3+1)- schemes) which are consistent with all the
experiments and the observations, despite the recent claims by the
Superkamiokande group that pure sterile oscillations
$\nu_e\leftrightarrow\nu_s$ in the solar neutrinos and pure sterile
oscillations $\nu_\mu\leftrightarrow\nu_s$ in the atmospheric
neutrinos are disfavored.  In particular, the reason that the (2+2)-scheme
is consistent with the recent 
Superkamiokande data is because both
solar and atmospheric neutrinos have hybrid oscillations
of active and sterile oscillations.

\section*{Acknowledgments}
I would like to thank Yoshitaka Kuno for
invitation and the local organizers for hospitality during the workshop.
This research was supported in
part by a Grant-in-Aid for Scientific Research of the Ministry of
Education, Science and Culture, \#12047222, \#10640280.

\end{document}